# Variations in the Solar Coronal Rotation with Altitude – Revisited

Hitaishi Bhatt[1*], Rupal Trivedi[2], Som Kumar Sharma[3], Hari Om Vats[3]

1 Dept. of Physics, M and N Virani Science College, Rajkot-360005, India.

2 Dept. of Physics, H. and H.B. Kotak Institute of Science, Rajkot-360004, India.

3 Physical Research Laboratory, Ahmadabad-380009, India.

∗Author for correspondence: bhatthitaishi@yahoo.com

## Abstract

Here we report in depth reanalysis of a paper by Vats *et al.* (2001) [*Astrophys. J.* 548, L87] based on the measurements of differential rotation with altitude as a function of observing frequencies (as lower and higher frequencies indicate higher and lower heights, respectively) in the solar corona. The radial differential rotation of the solar corona is estimated from daily measurements of the disc-integrated solar radio flux at 11 frequencies: (275, 405, 670, 810, 925, 1080, 1215, 1350, 1620, 1755 MHz and 2800 MHz). We use the same data as were used in Vats *et al.* (2001), but instead of the 12$^{th}$ maxima of autocorrelograms used there, we use the 1$^{st}$ secondary maxima to derive the synodic rotation period. We estimate synodic rotation by Gaussian fit of the 1$^{st}$ secondary maxima. Vats *et al.* (2001) reported that the sidereal rotation period increases with increasing frequency. The variation found by them was from 23.6 to 24.15 days in this frequency range with a difference of only 0.55 days. The present study finds that sidereal rotation period increases with decreasing frequency. The variation range is from 24.4 to 22.5 days and difference is about three times larger (1.9 days). However, at 925 MHz both studies give similar rotation period. In Vats *et al.* (2001) the Pearson's factor with trend line was 0.86 whereas present analysis obtained a ~0.97 Pearson's factor with the trend line. Our study shows that the solar corona rotates slower at higher altitudes, which is in contradiction to the findings reported in Vats *et al.* (2001).

Key words: Solar radio flux- flux modulation method- Gaussian fit- sidereal rotation period.

## 1. Introduction

The solar rotation is an important phenomenon and has caught the attention of many scientific research groups. The solar rotation, both in the interior of the Sun and in its atmosphere, have been extensively studied during last four decades. In the early days of solar research, the corona could be observed only during total solar eclipses. Presently, there are various ground-based (*i.e.* MAST, GONG, Cracow Astronomical Observatory) and space-based (*i.e. Yohkoh*, SDO, SOHO/EIT and LASCO) continuous observations of the Sun at different wavelengths which corresponds to different altitudes. These data are used to study various dynamical characteristics of the Sun as well as solar rotation in different regions of the Sun. There are mainly three methods for determining the rotation of the Sun, (i) by monitoring the motion of tracers such as, sunspots, faculae, plages, convection cells, oscillation wave patterns, low-level magnetic features, *etc.* (ii) by the Doppler shift of photospheric spectral lines, and (iii) studying emitted flux modulations.

Dupree and Henze (1972) used tracers from spectroheliograms, analysed the Lyman continuum and extreme ultraviolet (EUV) lines to determine the rotation rate of the solar chromosphere, transition region and corona. Nash *et al.* (1988) have proved that the rotation of coronal holes can be understood in terms of a current-free model of the coronal magnetic field, in which holes are the foot point locations of open field lines. Coronal X-ray bright points in soft X-ray filtergrams are used for the estimation of the solar rotation by Golub (1974). Furthermore, Golub (1974) has compared the rotation obtained from X-ray bright points with KPNO magnetograms from 1970-78. Rybăk (1994) used Fe xiv 5303 Å green images taken from coronographic network worldwide to obtain solar rotation period for the area of ± 30 deg. Brajša *et al.* (1999) have determined solar rotation with respect to altitudes (differential rotation with altitude) by using daily full-disk solar maps obtained at 37 GHz, the microwave low-brightness-temperature regions are traced for various phases of the solar cycle. Brajša *et al.* (2000) have determined solar rotational behaviour as well as north − south asymmetry from the radio emission at 37 GHz. Altrock (2003) has determined precisely the rotational behaviour of corona by using synoptic photoelectric observations of Fe xiv and Fe x. Altrock (2003) suggested that at the latitude of 60 deg and above the rotation is almost rigid in the rising phase of the solar activity while during low solar activity cycle the rotation has a pronounced differential component. Low and high brightness-temperature regions, (LTRs and HTRs) in the chromosphere at 37 GHz were traced by Brajša *et al.* (2009) to determine solar rotation period using full-disc solar radio maps. In order to determine the solar rotation period, Hara (2009) analysed X-ray bright points (XBPs) and also determined the differential rotation rate by using a tracer method. The coronal bright points of the green images are analysed by tracer method (segmentation algorithm). Small bright coronal structures in SOHO-EIT images for the time period of almost a solar cycle were traced by Wöhl *et al.* (2010) and found sidereal rotation period with latitude. Sudar *et al.* (2016) have used six months of SDO/AIA data for Fe xiv 5303 Å coronal emission line to determine the rotation period. By tracer method rotation has been compared for tracers in white light, Hα lines, Fe xiv green line, Fe x, soft X-rays, microwave, SOHO/EIT, EUV and magnetogram images.

The exploration of the rotation of the solar corona is relatively less mainly because of the following three reasons: (1) features are less distinct such as coronal holes, (2) corona cannot be observed easily because it is tenuous (very low density) compared to the other solar regions, and (3) it is very difficult to measure the magnetic field directly there [Zwaan (1987), Vats *et al.* (2001), Chandra (2010)]. The thickness of solar corona ranges from ~5000 to ~3.5 × $10^6$ km from the solar surface Chandra (2010).

In the past using different solar observational datasets by various investigators for the solar coronal rotation variations at various heights have been reported. Gowronska and Borkowski (1995) used 125 MHz daily solar radio flux data, Vats *et al.* (1998) and (2001) and Chandra (2010) used 2800 MHz daily solar radio flux data, Chandra, Vats and Iyer (2010) used SXT images, Chandra and Vats (2011) used NoRH images, Karachik *et al.* (2006) used coronal bright points from SOHO/EIT 19.1 nm images, Brajša *et al.* (2004) used 28.4 nm SOHO/EIT images coronal bright points, Weber *et al.* (1999) and Kariyappa (2008) used SXT images to obtain sidereal rotation period of corona.

The differential rotation as a function of height in the solar corona was reported for the first time by Vats *et al.* (2001). For radio emissions the height increases with decreasing frequency. Therefore, the observations at different frequencies give information at different altitudes in the solar corona. They found that the sidereal rotation period, at 2800 MHz, from the lower corona is ~24.1 days. There are two schemes for the estimation of solar coronal rotation using radio emission, (1) autocorrelation function method used by Vats *et al.* (2001), Chandra and Vats (2009), Chandra (2010), and Chandra and Vats (2011); (2) wavelet analysis method used by Temmer *et al.* (2006) and by Xie *et al.* (2012). In the present study, we have also used autocorrelation function method.

## 2. Observations

The continuous daily observations of disc-integrated solar radio flux measurements at 275, 405, 670, 810, 925, 1080, 1215, 1350, 1620, and 1755 MHz are taken from the Cracow Astronomical Observatory in Poland[1] and the measurements at 2800 MHz are from the Algonquin Radio Observatory in Canada[2]. The data sets for a period of 26 months (01 June, 1997 – 31 July, 1999) are used in the current study revisiting the same data sets used by Vats *et al.* (2001). A typical example of daily solar radio flux is plotted with the day numbers and shown in Figure 1. The sample plot shown in the figure is for the 1080 MHz solar radio flux. From Figure 1 the modulation due to solar rotation is clearly evident, but it is rather difficult to obtain accurate rotation period from this plot. Thus we further analysed the solar flux using autocorrelation analysis method.

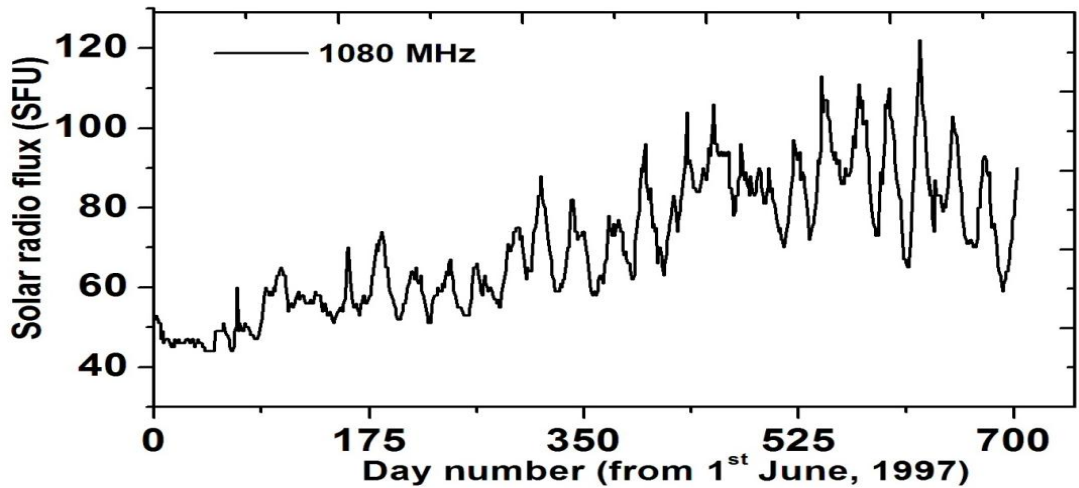

**Figure 1** Solar radio flux at 1080 MHz (solar flux unit) from 1st June 1997 to 31st July 1999 (703 days).

---

[1] http://www.oa.uj.edu.pl/sol/index.html
[2] http://www.ngdc.noaa.gov

## 3. Analysis and Results

The autocorrelation is defined by Equation (1) given below.

$$P_x(l) = P_x(-l) = \frac{\sum_{k=0}^{n-1}(x_k - \bar{x})(x_{k-1} - \bar{x})}{\sum_{k=0}^{n-1}(x_k - \bar{x})^2}, \quad (1)$$

where $l$ = Lag, n = No. of observations, $k$ = 0, 1, 2, 3, 4…., $P_x(l)$ = Auto correlation at lag l.

Data of all the 11 frequencies (given in the observations section) are analysed using above equation. The autocorrelograms obtained at three different frequencies are shown in Figure 2. Autocorrelograms shown in Figure 2 are nearly same as those obtained by Vats *et al.* (2001) [Figure 1c of Vats *et al.* (2001)]. Autocorrelograms show several maxima corresponding to synodic rotation period. For higher accuracy (≤0.1 day) 12$^{th}$ maximum was used by Vats *et al.* (2001). To estimate solar coronal rotation we use 1$^{st}$ secondary maximum. The 1$^{st}$ secondary maximum was also used by Chandra and Vats (2011), Xie *et al.* (2012) and by other researchers. The 1$^{st}$ secondary maxima were fitted by cosine fit by Chandra and Vats (2011) to obtain good estimation of the rotation period. In the present study Gaussian fit to the 1$^{st}$ secondary maxima is used with a view that the time corresponding to the peak of fit would provide a better estimate of the rotation period.

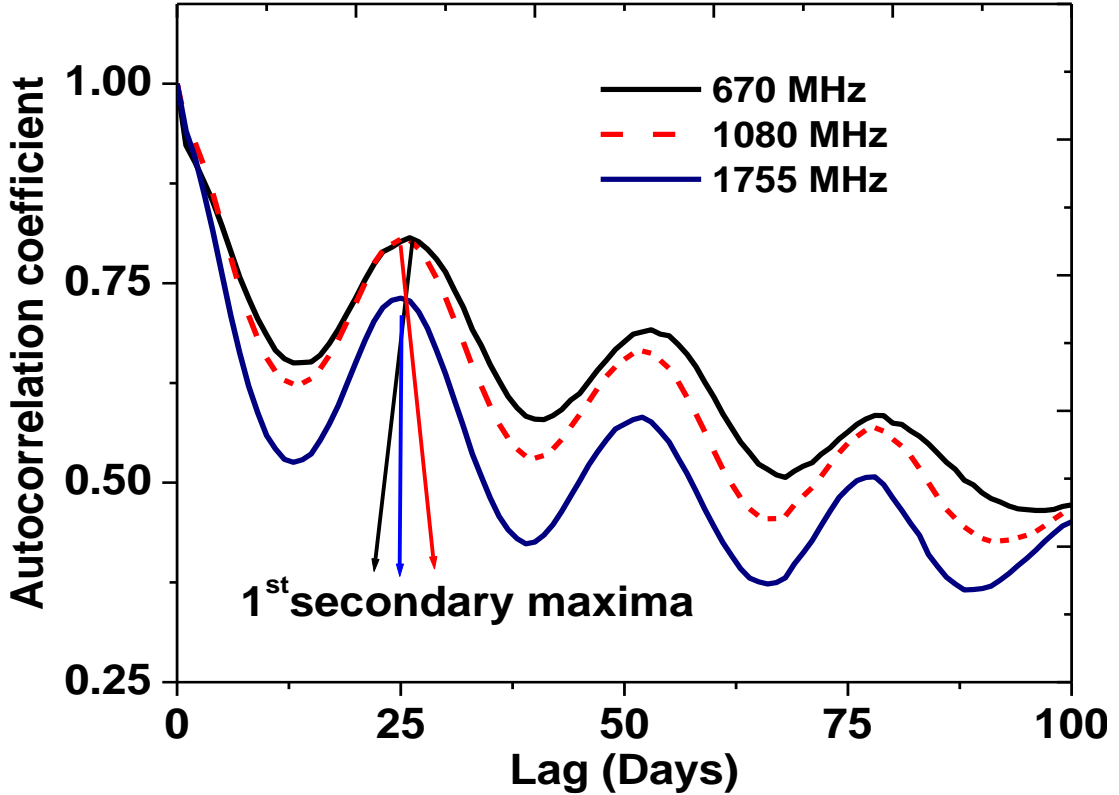

**Figure 2** Autocorrelograms for three different radio frequency fluxes as function of lag (days) for 100 lags which is represented as 670 MHz (Solid black line), 1080 MHz (Dotted red line) and 1755 MHz (Blue line) respectively.

In Figure 3 the Gaussian fit of the 1$^{st}$ secondary maxima is plotted for 1080 MHz observations and the centre of the Gaussian fit gives the synodic rotation period. The Gaussian fit is determined by

$$y = \frac{A}{w\sqrt{\frac{\pi}{2}}} e^{\frac{-2(x-x_0)^2}{w^2}} + y_0, \qquad (2)$$

where $x_o$= center of the maxima, $w = 2$ times the standard deviation of the Gaussian (2*$\sigma$) or approximately 0.849 times the width of the peak at half the peak value. $A$= area under the curve, $y_0$= baseline offset.

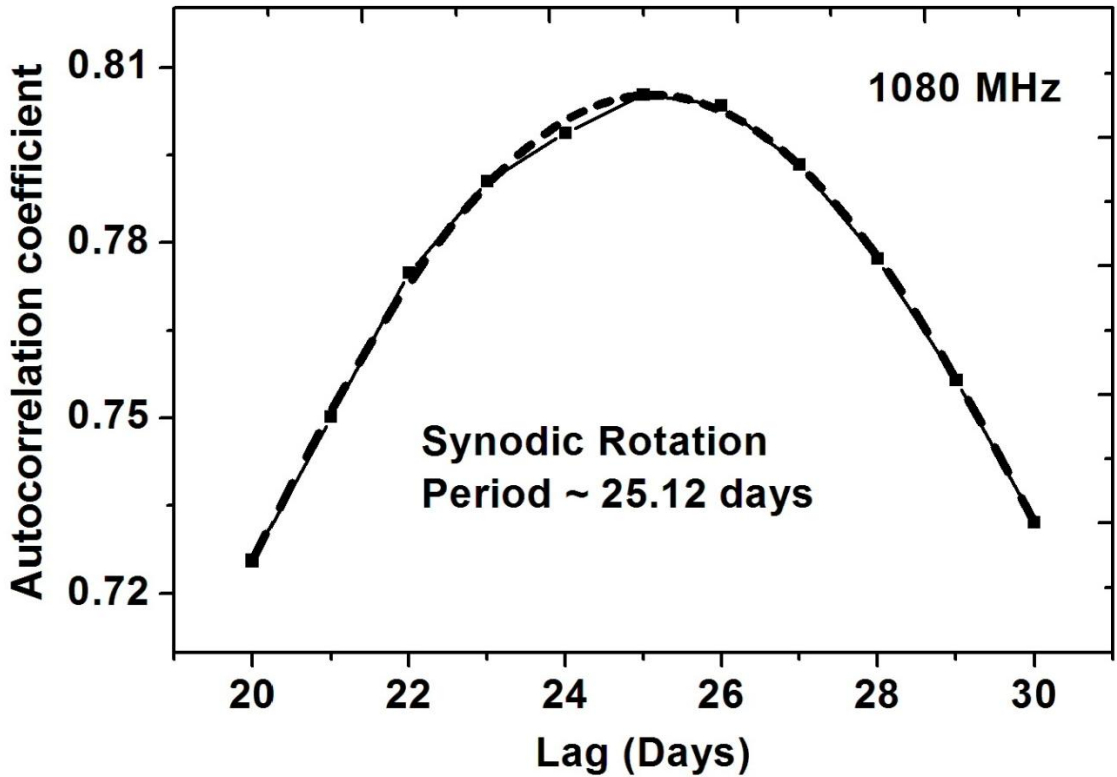

**Figure 3** Sample example of Gaussian fit over the 1$^{st}$ secondary maxima of the autocorrelogram at 1080 MHz (The dashed line represents the Gaussian fit and the solid line represents the 1$^{st}$ secondary maxima of autocorrelograms).

The synodic rotation period is the apparent rotation period of the Sun as seen from the Earth, which is orbiting around the Sun. Thus the sidereal rotation (actual) period can be determined using following relation

$$R = \frac{S \times 365.26}{S + 365.26}, \qquad (3)$$

where $R$ = sidereal period, $S$ = Synodic period.

The value of 365.26 is the number of days in an Earth sidereal year. The solar sidereal rotation period is estimated for only 9 radio frequencies, as the autocorrelation curves at 2 lower frequencies (275 and 405 MHz) are noisy. A plot of the sidereal rotation period as a function of solar radio frequencies of the present study as well as that by Vats *et al.* (2001) is shown in the Figure. 4. The trend line is a polynomial fit and which is fitted for the present study as well as for the Vats *et al.* (2001) results. The polynomial fit is evaluated by Polynomial function as defined by

$$p(x) = \sum_{k=1}^{j} a_k x^k, \qquad (4)$$

where k is the number of data points, *a* is constant

The polynomial fit is cross correlated and the correlation is characterized by the Pearson factor. The Pearson factor is defined as

$$R^2 = \frac{(x_1 - x_0)^2}{(y_1 - y_0)^2}, \qquad (5)$$

where $R$ is Pearson's factor, $x_1, y_1$ are data points and $x_0, y_0$ are base line points.

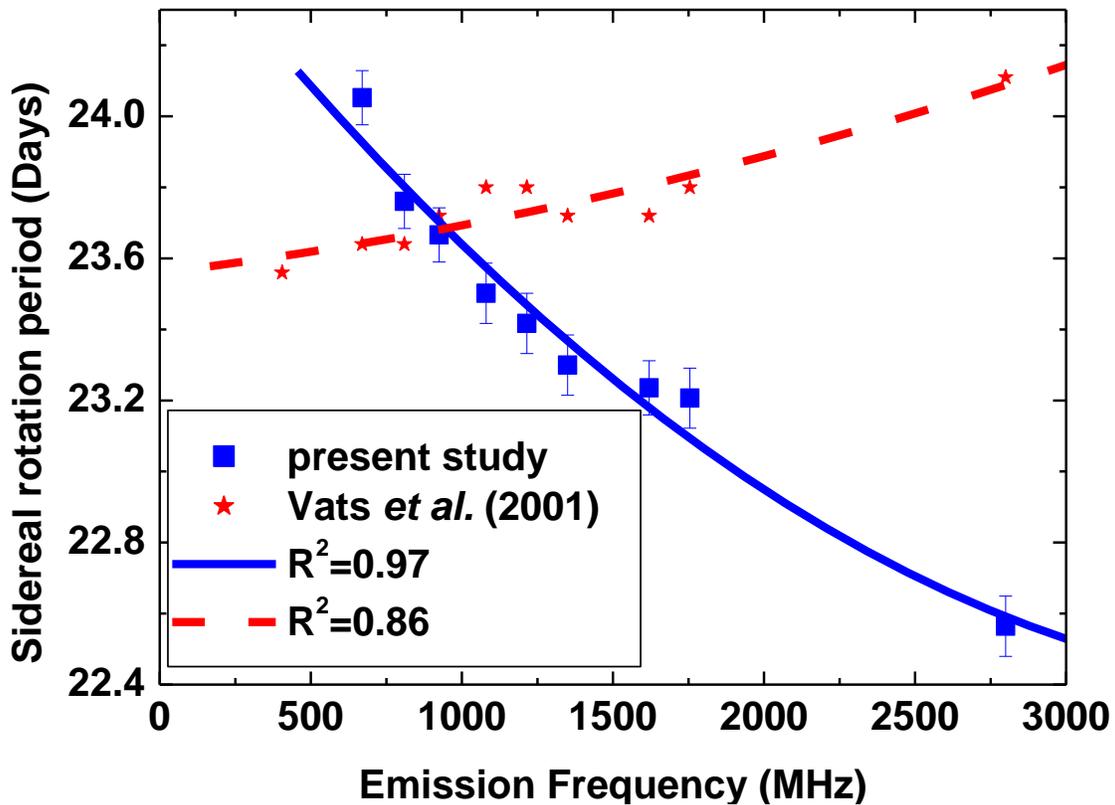

**Figure 4** Comparison of the results of the present study and those by Vats *et al.* (2001). The values of Vats *et al.* (2001) are shown by '*' and dashed red line is the polynomial fit. The values of the present study are shown by solid blue squares with error bars and the solid blue line is the polynomial fit.

Figure 4 shows a better fit with the trend line of the present study and hence the estimates are more reliable. The estimates of the present study show that the sidereal rotation period decreases with increasing frequency which in contrast to the findings of Vats *et al.* (2001). In the solar corona higher frequency radio emissions are emanating from the inner corona (lower heights), hence the present study reveals that the rotation period increases with altitude implying slower angular rotation at greater height. The variation of rotation period range from 24.4 to 22.5 days; the difference is ~1.9 days. The Pearson's factor of the polynomial $R^2$ is 0.97 which is very near to an ideal one. On the other hand the estimates of Vats *et al.* (2001) showed that the sidereal rotation period increases with increasing frequency or the rotation period decreases with increasing altitude. The variation of the sidereal rotation period ranges from 23.6 to 24.15 days; the difference is ~0.55 days. The Pearson's factor of the polynomial fit $R^2$ is 0.86 which is lower than that of the present study. For the frequency range of 670 MHz - 2800 MHz; present study shows angular velocity of the disk integrated solar corona range from 14.97 to 15.96 degrees per day whereas for Vats *et al.* (2001) it varies from 15.23 to 14.92 degrees per day.

## 4. Discussion

Here reanalysis of disk integrated radio flux at 11 frequencies are carried out for the estimation of solar coronal rotation period. Vats *et al.* (2001) used the same data and reported that solar coronal rotation period decrease with the increasing altitude in the solar corona. They have used $12^{th}$ maxima of autocorrelogram to estimate solar rotation period. In the present study $1^{st}$ secondary maxima is fitted with Gaussian curve and the peak of the fit is taken as proxy for the solar synodic rotation period. The estimates in the present study and Vats *et al.* (2001) are in qualitative agreement, however quantitatively significant difference have been noted. The present estimates show that the rotation period increases with increasing altitude. The variation is larger than those reported by Vats *et al.* (2001). The Pearson factor is found to be 0.97 in of the present study however it was 0.86 for Vats *et al.* (2001). It could be due to the fact that first peak is most significant and Gaussian fit to the peak would be better estimate of rotation period. Higher Pearson factor also indicate higher reliability of the present study. It is interesting to note that at 925 MHz both [present work and Vats *et al.* (2001)] give almost the same estimates. This could be due to low interference at 925 MHz which possibly increases for the frequencies lower and higher than 925 MHz. The interference may be due to scattering by the plasma irregularities. The plasma irregularities are usually there in the solar corona as well as in the Earth upper atmosphere. The irregularities of the Earth's upper atmosphere will have more effect on lower frequencies whereas coronal irregularities will have more effect on higher frequencies. It is likely that interference will effect higher order peaks much more than the first.

## 5. Conclusions

From Vats *et al.* (2001) the coronal rotation period decreases with altitude, the present reanalysis show that the coronal rotation period increases with altitude. However, the order of magnitude is nearly same in both. The reason for the opposite variation of coronal rotation velocity (period) with altitude is yet unknown. The interference is one possibility which may have more impact in the higher peaks of the autocorrelogram. Thus first secondary maximum as used in the present analysis is expected to give more accurate estimate of the rotation period. This would make the present method better than that of Vats *et al.* (2001) Moreover, an increase in period with altitude appears more logical as this would mean that outer corona lags behind inner corona. More research work is needed to further clarify this very interesting finding.

## Acknowledgement

The authors thank the referee for his very constructive suggestions to improve the quality of the manuscript. The data used in the study is the same as was used by Vats et al (2001). These were obtained from the two websites http://www.oa.uj.edu.pl/sol/index.html and http://www.ngdc.noaa.gov. The authors are thankful to Prof. K. N. Iyer and Dr. K. K. Shukla for their scientific suggestions to improve the quality of research content and presentation. This research work at PRL is supported by Department of Space, Govt. of India.

## Disclosure of Potential Conflicts of Interest

The authors declare that they have no conflicts of interest.